\documentclass[sigconf,10pt]{acmart}

\usepackage[acronyms]{glossaries}
\usepackage{soul}
\usepackage{siunitx} 
	\DeclareSIUnit\torr{Torr}
\usepackage{subfig}

\newcommand{\teranova}{TeraNova}

\newacronym{lidar}{LIDAR}{Light Detection and Ranging}
\newacronym{5g}{5G}{fifth generation}
\newacronym{mmwave}{mmWave}{millimeter wave}
\newacronym{phy}{PHY}{physical layer}
\newacronym{mac}{MAC}{medium access control}
\newacronym{uav}{UAV}{unmanned autonomous vehicle}
\newacronym{em}{EM}{electromagnetic}
\newacronym{iot}{IoT}{Internet of Things}
\newacronym{dl}{DL}{deep learning}
\newacronym{ml}{ML}{machine learning}
\newacronym{drl}{DRL}{deep reinforcement learning}
\newacronym{urc}{URC}{ultra-reliable computing}
\newacronym{urllc}{URLLC}{ultra-reliable low-latency communication}
\newacronym{mimo}{MIMO}{multiple-input multiple-output}
\newacronym{mu}{MU}{multi-user}
\newacronym{rfid}{RFID}{Radio Frequency Identification}
\newacronym{rfp}{RFP}{radio fingerprinting}
\newacronym{sdr}{SDR}{software-defined radio}
\newacronym{mas}{MAS}{Mobile Autonomous System}
\newacronym{rl}{RL}{reinforcement learning}
\newacronym{los}{LoS}{line-of-sight}
\newacronym{dnn}{DNN}{deep neural network}
\newacronym{fpga}{FPGA}{field-programmable gate array}
\newacronym{cv}{CV}{computer vision}
\newacronym{mcs}{MCS}{modulation and coding scheme}
\newacronym{soc}{SoC}{system-on-chip}
\newacronym{mumimo}{MU-MIMO}{\gls{mu}-\gls{mimo}}
\newacronym{dsp}{DSP}{digital signal processing}
\newacronym{snr}{SNR}{signal-to-noise ratio}
\newacronym{csi}{CSI}{channel state information}
\newacronym{svd}{SVD}{singular value decomposition}
\newacronym{cs}{CS}{compressive sensing}
\newacronym{ism}{ISM}{industrial, scientific and medical}
\newacronym{dsa}{DSA}{dynamic spectrum access}
\newacronym{cnn}{CNN}{convolutional neural network}
\newacronym{rfsoc}{RFSoC}{\gls{rf} System on Chip}
\newacronym{ntp}{NTP}{network time protocol}
\newacronym{ptp}{PTP}{precise time protocol}
\newacronym{nu}{NU}{Northeastern University}
\newacronym{tamu}{TAMU}{Texas A\&M University}
\newacronym{ncsu}{NCSU}{North Carolina State University}
\newacronym{uta}{UTA}{University of Texas Austin}
\newacronym{fiu}{FIU}{Florida International University}
\newacronym{uo}{OU}{University of Oklahoma}
\newacronym{gt}{GT}{Georgia Tech}
\newacronym{ucb}{UCB}{University of California Berkeley}
\newacronym{ucsb}{UCSB}{University of California Santa Barbara}
\newacronym{ttu}{TTU}{Texas Tech University}
\newacronym{uh}{UH}{University of Hawaii}
\newacronym{eess}{EESS}{Earth exploration satellite services}
\newacronym{nsf}{NSF}{National Science Foundation}
\newacronym{ntia}{NTIA}{National Telecommunications and Information Administration}
\newacronym{rfc}{RFC}{Request for Comments}
\newacronym{dod}{DoD}{Department of Defense}
\newacronym{frp}{FRP}{foundational research principle}
\newacronym{cct}{CCT}{technological cross-cutting theme}
\newacronym{wiot}{WIoT}{Institute for the Wireless Internet of Things}
\newacronym{ewd}{EWD}{education and workforce development}
\newacronym{pawr}{PAWR}{Platforms for Advanced Wireless Research}
\newacronym{ppo}{PPO}{\gls{pawr} Project Office}
\newacronym{iq}{IQ}{in-phase and quadrature}
\newacronym{if}{IF}{intermediate frequency}
\newacronym{pl}{PL}{physical logic}
\newacronym{lna}{LNA}{low-noise amplifier}
\newacronym{rf}{RF}{radio frequency}
\newacronym{wpan}{WPAN}{wireless personal area network}
\newacronym{wlan}{WLAN}{wireless local area network}
\newacronym{wan}{WAN}{Wide Area Network}
\newacronym{6g}{6G}{sixth generation}
\newacronym{vr}{VR}{virtual reality}
\newacronym{ar}{AR}{augmented reality}
\newacronym{src}{SRC}{Semiconductor Research Corporation}
\newacronym{darpa}{DARPA}{Defense Advanced Research Projects Agency}
\newacronym{adc}{ADC}{analog to digital converter}
\newacronym{dac}{DAC}{digital to analog converter}
\newacronym{di}{DI}{deionized}
\newacronym{gsaps}{GSaps}{Giga-samples-per-second}
\newacronym{awg}{AWG}{arbitrary waveform generator}
\newacronym{dso}{DSO}{digital storage oscilloscope}
\newacronym{nlos}{NLoS}{non-line-of-sight}
\newacronym{thz}{THz}{terahertz}
\newacronym{ghz}{GHz}{gigahertz}
\newacronym{si}{Si}{silicon}
\newacronym{soi}{SoI}{Silicon-on-Insulator}
\newacronym{sige}{SiGe}{Silicon-Germanium}
\newacronym{inp}{InP}{Indium Phosphide}
\newacronym{gan}{GaN}{Gallium Nitride}
\newacronym{gaas}{GaAs}{Gallium Arsenide}
\newacronym{jpl}{JPL}{Jet Propulsion Laboratory}
\newacronym{ic}{IC}{integrated circuit}
\newacronym{ipa}{IPA}{isopropyl alcohol}
\newacronym{hbt}{HBT}{heterojunction bipolar transistor}
\newacronym{hemt}{HEMT}{high-electron mobility transistor}
\newacronym{pa}{PA}{power amplifier}
\newacronym{hdl}{HDL}{hardware description language}
\newacronym{fft}{FFT}{fast Fourier transform}
\newacronym{css}{CSS}{chirp spread spectrum}
\newacronym{dsss}{DSSS}{direct-sequence spread spectrum}
\newacronym{rssi}{RSSI}{received signal strength indicator}
\newacronym{bs}{BS}{base station}
\newacronym{ue}{UE}{user equipment}
\newacronym{nrdz}{NRDZ}{National Radio Dynamic Zone}
\newacronym{ofdm}{OFDM}{Orthogonal Frequency Division Multiplexing}
\newacronym{leo}{LEO}{low Earth orbiting}
\newacronym{trl}{TRL}{technology readiness level}
\newacronym{ldgm}{LDGM}{low-density generator matrix}
\newacronym{ldpc}{LDPC}{low-density parity-check}
\newacronym{lo}{LO}{local oscillator}
\newacronym{isec}{ISEC}{Interdisciplinary Science and Engineering Complex}

\newacronym{osa}{OSA}{OpenAirInterface Software Alliance}
\newacronym{casper}{CASPER}{Collaboration for Astronomy Signal Processing and Electronics Research}
\newacronym{qos}{QoS}{Quality of Service}
\newacronym{oran}{O-RAN}{Open Radio Access Network}
\newacronym{ran}{RAN}{Radio Access Network}

\newacronym{ric}{RIC}{RAN Intelligent Controller}
\newacronym{cbrs}{CBRS}{Citizens Broadband Radio Service}
\newacronym{gaa}{GAA}{General Authorized Access}
\newacronym{pal}{PAL}{Priority Access Licensee}
\newacronym{fcc}{FCC}{Federal Communications Commission}
\newacronym{sas}{SAS}{spectrum access system}
\newacronym{ai}{AI}{artificial intelligence}

\newacronym{ser}{SER}{symbol error rate}
\newacronym{rfic}{RFIC}{radio frequency integrated circuit}
\newacronym{rfi}{RFI}{\gls{rf} Interference}
\newacronym{aml}{AML}{adversarial machine learning}
\newacronym{sdn}{SDN}{software-defined networking}
\newacronym{star}{STAR}{Simultaneous Transmit and Receive}
\newacronym{sinr}{SINR}{signal-to-interference-noise ratio}
\newacronym{vlba}{VLBA}{Very Long Baseline Array}
\newacronym{ngvla}{ngVLA}{Next Generation Very Large Array}
\newacronym{nrao}{NRAO}{National Radio Astronomy Observatory}
\newacronym{fso}{FSO}{Federated Spectrum Observatory}
\newacronym{ngso}{NGSO}{non-geostationary orbit}
\newacronym{vsd}{VSD}{Value-Sensitive Design}
\newacronym{sensr}{SENSR}{Spectrum Efficient National Surveillance Radar}
\newacronym{gbps}{Gbps}{gigabits per second}
\newacronym{tbps}{Tbps}{Terabit-per-second}
\newacronym{nas}{NAS}{Network Attached Storage}
\newacronym{5gb}{5GB}{5G-and-beyond}
\newacronym{osi}{OSI}{Open Systems Interconnection model}
\newacronym{onr}{ONR}{Office of Naval Research}
\newacronym{afosr}{AFOSR}{Air Force Office of Scientific Research}
\newacronym{afrl}{AFRL}{Air Force Research Laboratory}
\newacronym{arl}{ARL}{Army Research Laboratory}
\newacronym{bdss}{BDSS}{broadband directional spectrum sensor}

\newacronym{aoa}{AoA}{Angle of Arrival}
\newacronym{noe}{NOE}{NRDZ Orchestration Engine}

\newacronym{mchem}{MCHEM}{Massive Channel Emulator}
\newacronym{afc}{AFC}{Automated Frequency Coordination}
\newacronym{esc}{ESC}{Environmental Sensing Capability}

\newacronym[firstplural=Devices Under Test (DUTs)]{dut}{DUT}{Device Under Test}

\newacronym{kpi}{KPI}{Key Performance Indicator}
\newacronym{dei}{DEI}{diversity, equity and inclusion}

\newacronym{itu}{ITU}{International Telecommunication Union}
\newacronym[firstplural=Notices of Inquiry (NOIs)]{noi}{NOI}{Notice of Inquiry}
\newacronym{wp}{WP}{Working Party}
\newacronym{drs}{DRS}{Digital Repository Service}
\newacronym{ms}{M.S.}{Master of Science}
\newacronym{apsk}{APSK}{Amplitude and Phase Shift Keying}
\newacronym{hbm}{HBM}{Hierarchical Bandwidth Modulation}
\newacronym{lme}{LME}{Load Modulation Effects}
\newacronym{eirp}{EIRP}{Effective Isotropic Radiated Power}
\newacronym{cte}{CTE}{Continuous-Time Equalizer}

\newacronym{ofdma}{OFDMA}{Orthogonal Frequency Division Multiplexing Access}
\newacronym{otfs}{OTFS}{Orthogonal Time Frequency Space}
\newacronym{papr}{PAPR}{Peak-to-Average Power Ratio}
\newacronym{mf}{MF}{Merit Factor}
\newacronym{mmse}{MMSE}{Minimum Mean Squared Error}
\newacronym{cnt}{CNT}{Carbon Nanotube}
\newacronym{ris}{RIS}{reconfigurable intelligent surface}
\newacronym{xr}{XR}{extended reality}
\newacronym{irs}{IRS}{intelligent reflecting surface}
\newacronym{ap}{AP}{access point}
\newacronym{cw}{CW}{continuous-wave}
\newacronym{thz-tds}{THz-TDS}{terahertz time-domain spectroscopy}
\newacronym{ber}{BER}{bit error rate}
\newacronym{ir}{IR}{infrared}
\newacronym{psg}{PSG}{programmable signal generator}
\newacronym{dsb}{DSB}{double sideband}
\newacronym{vdi}{VDI}{Virginia Diodes, Inc.}
\newacronym{awgn}{AWGN}{additive white Gaussian noise}

\AtBeginDocument{%
  }

\acmYear{2023}\copyrightyear{2023}
\setcopyright{acmlicensed}
\acmConference[mmNets '23]{The 7th ACM Workshop on Millimeter-Wave and Terahertz Networks and Sensing Systems}{October 6, 2023}{Madrid, Spain}
\acmBooktitle{The 7th ACM Workshop on Millimeter-Wave and Terahertz Networks and Sensing Systems (mmNets '23), October 6, 2023, Madrid, Spain}
\acmPrice{15.00}

\begin{document}

\title[Design and Validation of a Metallic Reflectarray for Communications at True THz Frequencies]{Design and Validation of a Metallic Reflectarray for Communications at True Terahertz Frequencies}


\author{Sherif Badran}
\authornote{Corresponding author.}
\orcid{0000-0002-4345-6329}
\affiliation{
  \institution{Northeastern University}
  \city{Boston}
  \state{Massachusetts}
  \country{USA}
}
\email{badran.s@northeastern.edu}

\author{Arjun Singh}
\orcid{0000-0003-0698-6790}
\affiliation{
  \institution{SUNY Polytechnic Institute}
  \city{Utica}
  \state{New York}
  \country{USA}
}
\email{singha8@sunypoly.edu}

\author{Arpit Jaiswal}
\orcid{0000-0002-4421-8816}
\affiliation{
  \institution{University at Buffalo}
  \city{Buffalo}
  \state{New York}
  \country{USA}
}
\email{arpitjai@buffalo.edu}

\author{Erik Einarsson}
\orcid{0000-0002-3896-2673}
\affiliation{
  \institution{University at Buffalo}
  \city{Buffalo}
  \state{New York}
  \country{USA}
}
\email{erikeina@buffalo.edu}

\author{Josep M. Jornet}
\orcid{0000-0001-6351-1754}
\affiliation{
  \institution{Northeastern University}
  \city{Boston}
  \state{Massachusetts}
  \country{USA}
}
\email{j.jornet@northeastern.edu}

\renewcommand{\shortauthors}{Badran et al.}

\begin{abstract}
    Wireless communications in the terahertz band (\SIrange{0.1}{10}{\tera\hertz}) is a promising and key wireless technology enabling ultra-high data rate communication over multi-gigahertz-wide bandwidths, thus fulfilling the demand for denser networks. The complex propagation environment at such high frequencies introduces several challenges, such as high spreading and molecular absorption losses. As such, intelligent reflecting surfaces have been proposed as a promising solution to enable communication in the presence of blockage or to aid a resource-limited quasi-omnidirectional transmitter direct its radiated power. In this paper, we present a metallic reflect\-array design achieving controlled non-specular reflection at true terahertz frequencies (i.e., \SIrange{1}{1.05}{\tera\hertz}). We conduct extensive experiments to further characterize and validate its working principle using terahertz time-domain spectroscopy and demonstrate its effectiveness with information-carrying signals using a continuous-wave terahertz testbed. Our results show that the reflectarray can help facilitate robust communication links over non-specular paths and improve the reliability of terahertz communications, thereby unleashing the true potential of the terahertz band.
\end{abstract}





 
\maketitle

\glsresetall

\section{Introduction}
\label{sec:intro}

The \gls{thz} band has been envisioned as a key asset for the next generation of wireless communication and sensing systems~\cite{akyildiz2022terahertz}. The very large bandwidth available at \gls{thz} frequencies (easily tens to hundreds of gigahertz) opens the door to terabit-per-second links in front-haul applications, as in wireless immersive \gls{xr}, and 
up to hundred \gls{gbps} in wireless backhaul applications, which can bridge the rural digital divide. In addition, the shorter wavelength of \gls{thz} signals opens the door to sub-millimetric resolution in radar applications. Moreover, the shorter wavelength also leads to smaller antennas that can be leveraged in nanonetworking applications~\cite{jornet2023nanonetworking}, including wireless networks-on-chip or wireless nano-bio sensor and actuator networks. Finally, the low but non-negligible photon energy of sub-terahertz and terahertz waves---from \SI{0.41}{\milli\electronvolt} up to \SI{41}{\milli\electronvolt}---leads to very unique \gls{em} signatures in different elements (from atmospheric gases to nano and biomaterials), 
which can be leveraged, for example, for target classification purposes.

All these opportunities come with a cost. First, the molecular absorption losses due to water vapor 
practically divide the \gls{thz} band into multiple absorption-defined windows, where the usable bandwidth changes with distance as well as other medium parameters. 
Second, the very small size of \gls{thz} antennas leads to a low effective area. While this is not a problem for nanonetworking applications, as the expected transmission range is usually under one meter, the small effective area of the antenna leads to high spreading losses for macroscale applications of the \gls{thz} band. This requires the adoption of high-gain directional antennas and/or focusing lenses. Finally, the interaction of \gls{thz} radiation with not just gases but effectively most materials can lead to significant blockage resulting from signal absorption and/or reflection.


To overcome the challenging propagation of \gls{thz} signals, the adoption of \glspl{irs} has been proposed~\cite{di2020smart, renzo2019smart,niu2013experimental}. \glspl{irs} can engineer the reflection of \gls{em} signals, introducing, for example, non-specular reflections, as well as more advanced functionalities such as polarization switching or
wavefront engineering, including the transformation of spherical or Gaussian wavefronts into more robust Bessel beams~\cite{arjun_wcm_2023}. Physically, \glspl{irs} come mostly in two forms, namely, reflectarrays and metasurfaces. Reflect\-arrays 
are integrated by metallic reflecting elements whose size and spacing are on the order of half a wavelength. The reflection phase or delay associated with each element can be set by means of controllable delay lines~\cite{nayeri2018reflectarray}. 
Metasurfaces, instead, are integrated by elements that are much smaller than the \gls{em} signal wavelength, known as meta-atoms. The smaller element size leads to enhanced functionalities but also results in more challenging control of the elements. 

Although the end goal is usually to have tunable reflect\-arrays and metasurfaces, tunability is not always needed. In many contexts, having a fixed response is sufficient, such as in the case of an indoor communication scenario where there are fixed or stationary blockers (i.e., walls, pillars, etc.)\@ In such cases, mounting, for instance, a reflectarray to steer an incoming beam in a fixed direction to another repeater or \gls{ap} is sufficient. Similarly, in nanoscale applications, reflections from the chip surface can be preprogrammed to provide connectivity with different cores~\cite{abadal2019wave}. Removing the reconfigurability requirement of \glspl{irs} drastically simplifies their design, fabrication, and control. This is particularly true at higher operation frequencies, where common phase/delay control elements are not available~\cite{singh2020d}.

Towards this goal, in this paper, we design, build, and experimentally characterize a preprogrammed reflectarray that operates in the first absorption-defined window above \SI{1}{\tera\hertz}, i.e., between 1 and \SI{1.05}{\tera\hertz}. The proposed design consists of an array of metallic reflecting patches with delay-controlling metallic stubs with micrometric dimensions that are fabricated with dimensions tailored to the specific criteria (Sec.~\ref{sec:design}). 
We experimentally characterize the structure using two complementary approaches, namely, 
broadband \gls{thz-tds} and narrowband communication using data-bearing \gls{cw} signals (Sec.~\ref{sec:exp validation}). With the results at hand, we then outline the next steps and potential future directions (Sec.~\ref{sec:conclusion}).


\section{Design and Fabrication}
\label{sec:design}

In this section, we detail the design of the individual metallic patch, as well as the resulting reflectarray. We then explain the steps involved in the fabrication of the same. 

\subsection{Reflectarray Design and Principle}
\subsubsection{Patch}

The fundamental radiating element of the reflectarray is a metallic patch. As per~\cite{balanis2015antenna}, we utilized the cavity model to derive the required width $W$ and the length $L$ of the patch at a given design frequency $f_{0}$: 
\begin{equation}
\label{eq:width}
    W = \frac{c}{2f_{0}\sqrt{\frac{\varepsilon_{r}+1}{2}}}, 
\end{equation}
\begin{equation}
\label{eq:length}
    L = \frac{c}{2f_{0}\sqrt{\varepsilon_\mathrm{eff}}} - 0.824h\biggl[\frac{(\varepsilon_\mathrm{eff}+0.3)(\frac{W}{h}+0.264)}{(\varepsilon_\mathrm{eff}-0.258)(\frac{W}{h}+0.8)}\biggr].
\end{equation}
Here, $h$ represents the thickness of the substrate, with $\varepsilon_{r}$ denoting the dielectric constant. As specified in~\cite{balanis2015antenna}, $h$ should be in the range of \numrange{0.003}{0.05}~$\lambda_{0}$, where $\lambda_{0} = c/f_{0}$. Due to the fringing effect at the edges of the patch, the effective dielectric constant $\varepsilon_\mathrm{eff}$, is given by~\cite{balanis2015antenna}:
\begin{equation}
\label{eq:var_epsilon}
    \varepsilon_\mathrm{eff} = \frac{\varepsilon_{r}+1}{2} + \frac{\varepsilon_{r}-1}{2}[1+12h/W]^{-1/2}.
\end{equation} 

The patch is first designed in transmission mode, where the $S_{11}$ parameter, or the reflection coefficient~\cite{balanis2015antenna}, is utilized to ensure resonance at the design frequency. By the symmetry of \gls{em} operation, a good transmitter is also a good receiver, and thus, the patch is a good reflector as it can both receive and reradiate \gls{em} waves. Since the patch is vertically polarized, the design works for linearly polarized waves~\cite{nayeri2018reflectarray}.


\begin{figure*}
	\centering
      \subfloat[Optical micrograph of reflectarray (scale is \SI{100}{\micro\meter}) \label{fig:array-micro}]{\includegraphics[width=0.88\columnwidth]{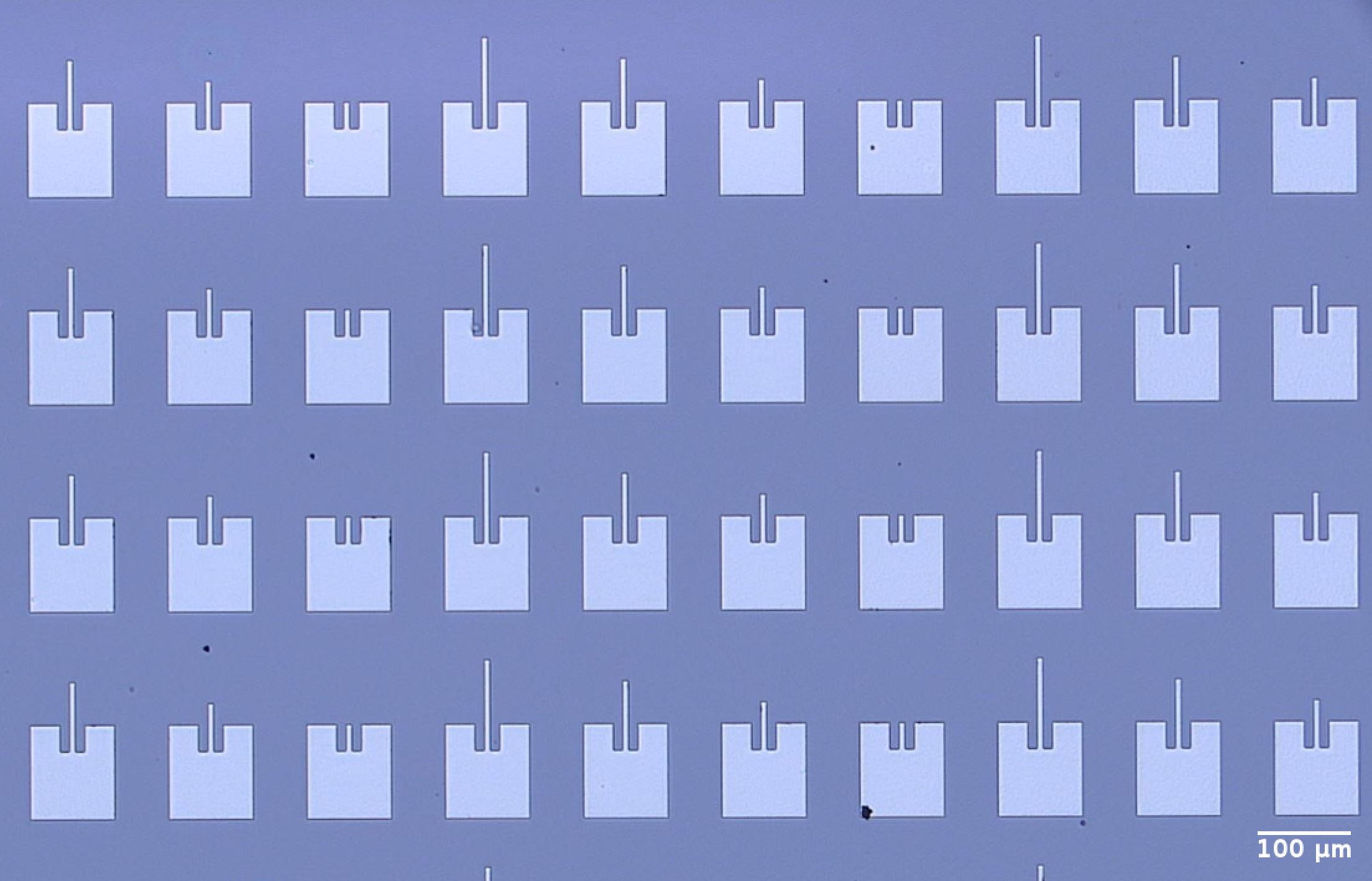}}
    \hfill
    \subfloat[Cross-sectional schematic of element structure \label{fig:fab-stack}]{\includegraphics[width=0.75\columnwidth]{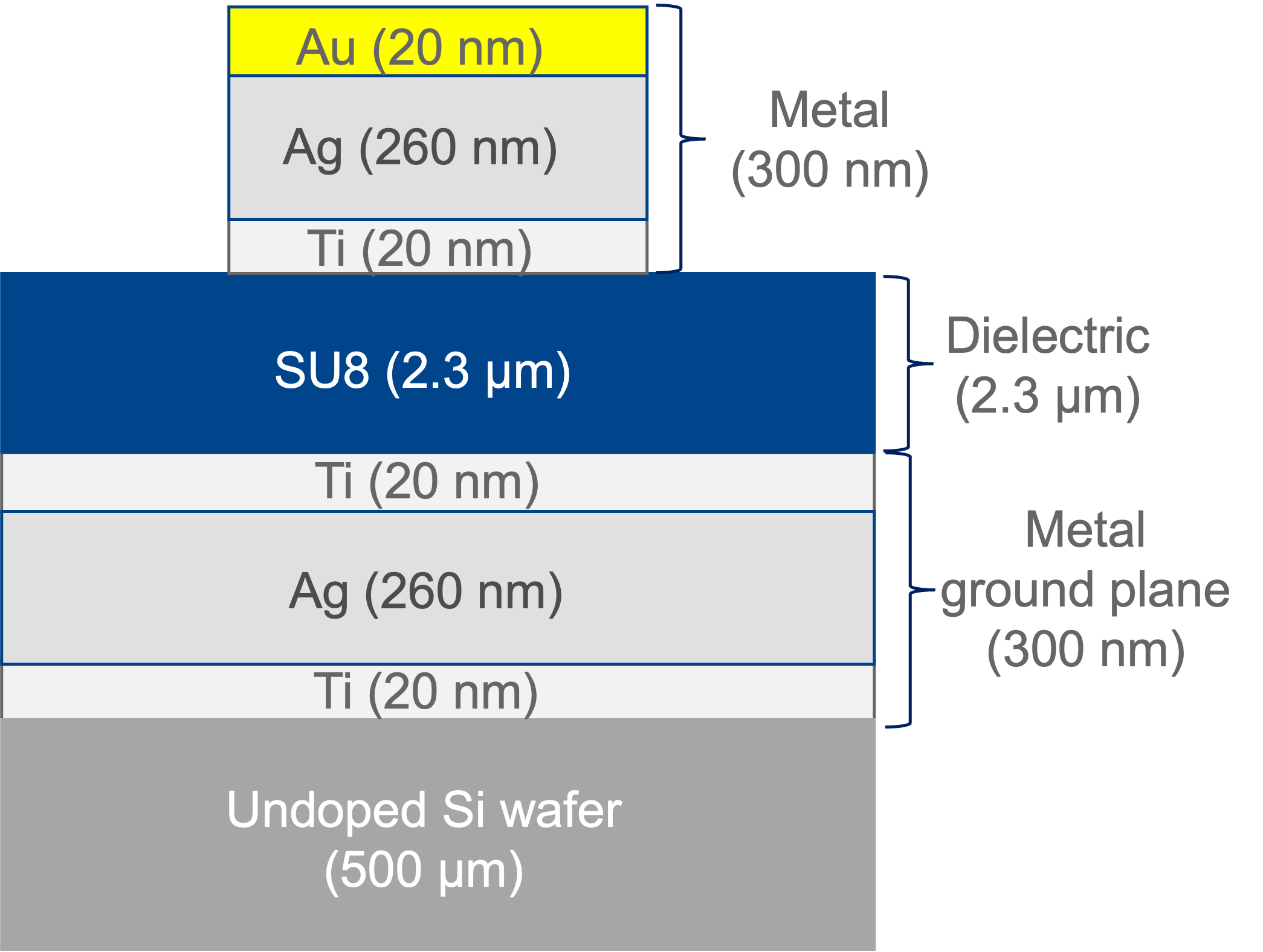}}
    \hfill
     \subfloat[Photo of reflectarray \label{fig:irs-sample}]
    {\includegraphics[width=0.4\columnwidth]{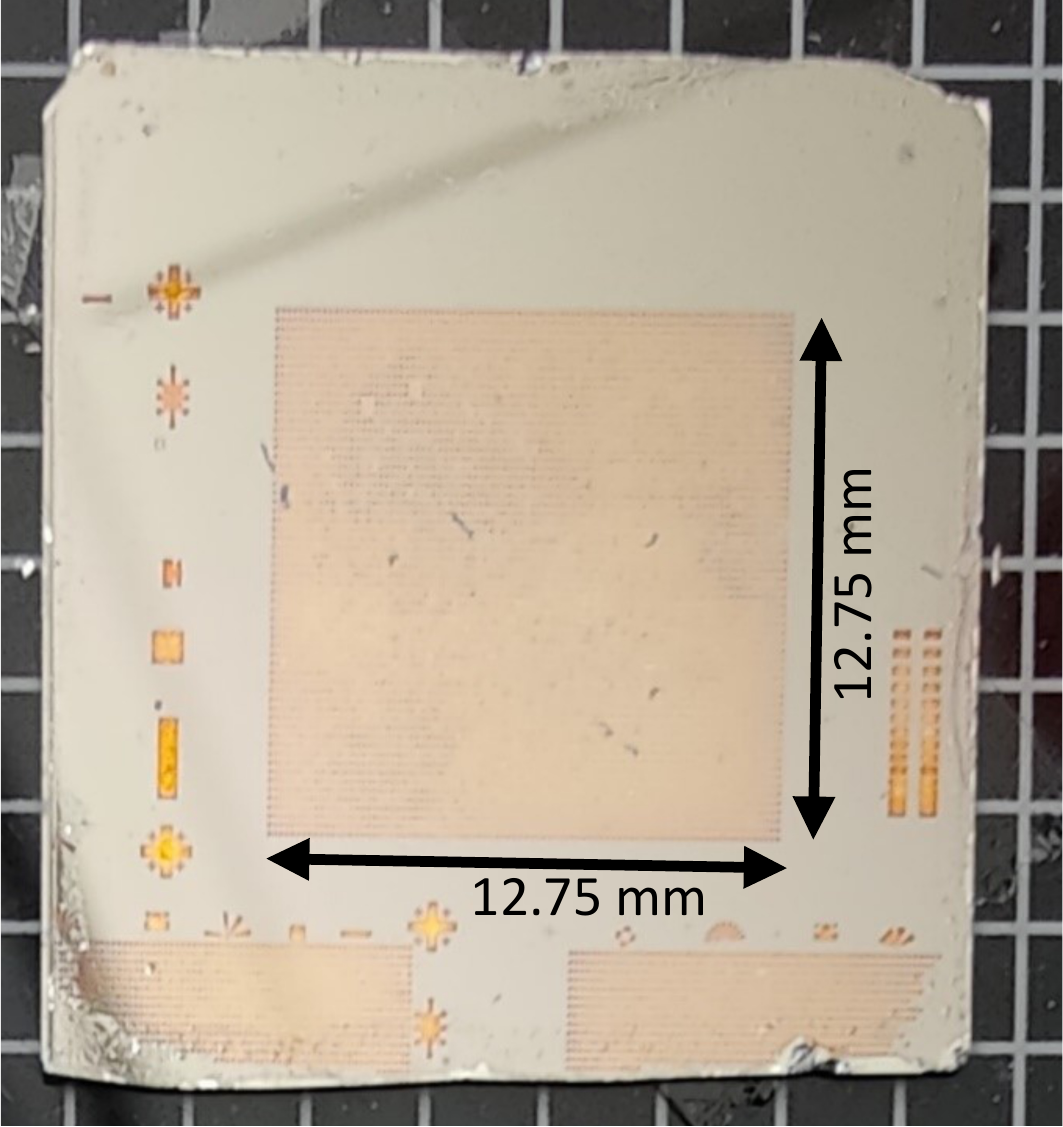}}
    \caption{Schematic and optical images of the reflectarray. The size of the patterned area in (c) is \num{12.75} $\times$ \SI{12.75}{\milli\meter}.}
	\label{fig:array-images}
\end{figure*}

\subsubsection{Reflectarray Integration}
Following common reflect\-array design principles, we designed the individual patches to be separated by a center-to-center distance of $0.5\lambda_{0}$. Further, we recall from array theory that to direct a beam in an angle $\theta$, the progressive phase delay $\Phi_{RA}$ across the reflectarray elements should be:
\begin{equation}
\label{eq:phase_delay}
    \Phi_{RA} = k_{0}(R_{i} - d\sin\theta), 
\end{equation} 
where $k_{0}$ is the free-space wave vector and $d$ is the distance between the elements. $R_{i}$ is the additional phase that comes from the direction of incidence, and in the case of broadside radiation, this term vanishes, yielding the familiar $\Phi_{RA} = -k_{0}d\sin\theta$ term. As the phase delay is relative, we derived a pattern that can be easily replicated across a large array. Namely, we chose to implement a $\pi/2$ radians progressive phase shift, which wraps around $2\pi$ every four elements. The resultant direction of steering is then \ang{30} as per~\eqref{eq:phase_delay}. The delay is implemented through a fixed-length delay stub along the resonant length of the patch. The length is calculated as per the phase constant. Namely, to effect a delay of $\Phi$ radians, the length of the stub $L$ should be $L = \Phi / k$, 
where $k$ is the wave vector within the substrate. Based on these principles, we designed the final reflectarray as shown in Fig.~\ref{fig:array-micro}, where the pattern can be seen to repeat every four elements.

\subsection{Fabrication}
\label{sec:fabrication}



We fabricated the reflectarray on a substrate consisting of a \SI{300}{\nm} (\SI{20}{\nm} Ti + \SI{260}{\nm} Ag + \SI{20}{\nm} Ti) metallic ground plane deposited atop a Si/SiO\textsubscript{2} wafer
to ensure high reflectivity~\cite{abohmra2020terahertz}. 
A \SI{2.3}{\micro\meter} SU8 layer was spin-coated on the ground plane and the reflectarray designs were patterned atop the SU8 dielectric layer using conventional photolithography.

The Si/SiO\textsubscript{2} substrate was cleaned by sonication in acetone, isopropyl alcohol, and \gls{di} water for \SI{5}{\minute} each, followed by N\textsubscript{2} blow dry and heating for \SI{5}{\minute} at \SI{150}{\degreeCelsius}. 
The substrate was then treated with O\textsubscript{2} plasma at \SI{65}{\watt} for \SI{120}{\second} 
to remove organic contaminants. 
The cleaned substrate was then inserted into an electron beam evaporation system 
for deposition of the ground plane metals. 
The structure of the deposited metal (see Fig.~\ref{fig:fab-stack}) was \SI{20}{\nm} Ti (for adhesion), \SI{260}{\nm} Ag (for reflection), and another \SI{20}{\nm} Ti adhesion layer. 
%
%
%
Photolithography was used to write the array pattern onto the substrate. For improved lift-off, LOR3B photoresist was first spin coated 
at \SI{2500}{rpm} for \SI{45}{\second}, followed by a post-coat bake at \SI{190}{\degreeCelsius} for \SI{4}{\minute}. A second photoresist (Microposit S1813) was then spin coated at \SI{5000}{rpm} for \SI{45}{\second} and followed by post-coat bake at \SI{115}{\degreeCelsius} for \SI{1}{\minute}. 
The substrate was subsequently placed in contact with the designed photomask 
and exposed (Hg i-line, \SI{350}{\watt}, \SI{4.5}{\ampere}) for \SI{50}{\second}. 

After photolithography, the photoresists were developed in Microposit MF-319 solution for \SI{45}{\second} followed by rinsing in \gls{di} water for \SI{30}{\second}. 
A similar \SI{300}{\nm} metal stack (\SI{20}{\nano\meter} Ti, \SI{260}{\nm} Ag, and \SI{20}{\nm} Au) was deposited to form the metal arrays. 
The topmost Au layer was deposited to prevent the Ag surface from oxidation. The sample was then left in Microposit Remover 1165 solution overnight at \SI{70}{\degreeCelsius}. After lift-off, the sample was rinsed in \gls{ipa} and \gls{di} water for \SI{5}{\minute} each, followed by N\textsubscript{2} blow dry.
A photo of the resulting reflectarray is shown in Fig.~\ref{fig:irs-sample}.

\section{Experimental Validation}
\label{sec:exp validation}

In this section, we validate the reflectarray working principle using two platforms, i.e., \gls{thz-tds} and a \gls{cw} \gls{thz} testbed.

\subsection{Terahertz Time-Domain Spectroscopy}
\begin{figure}
    \centering
    \includegraphics[width=\columnwidth]{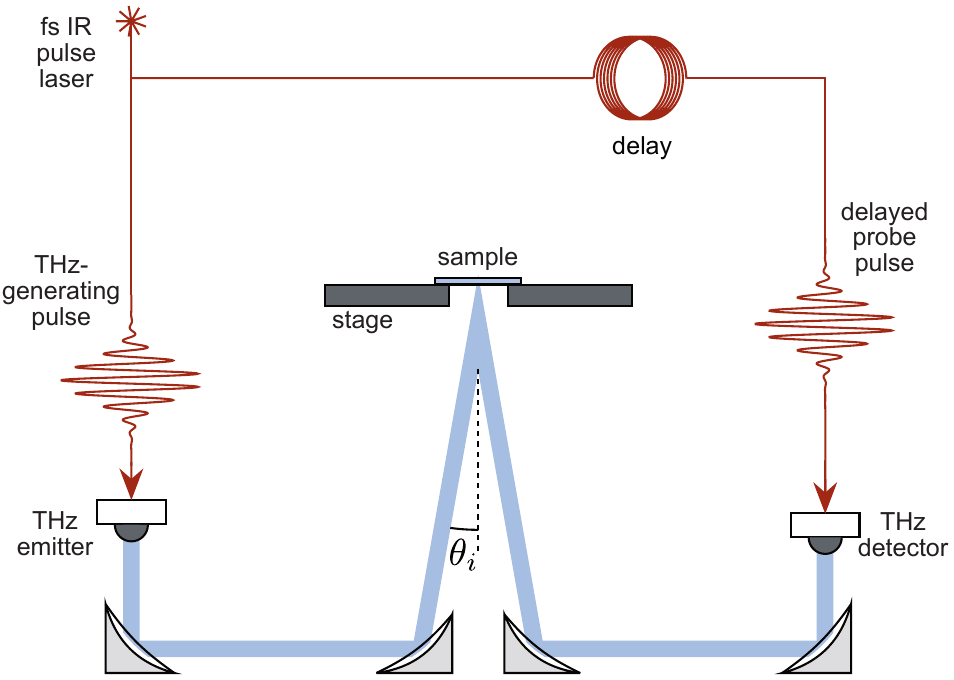}
    \caption{Schematic of the \gls{thz-tds} measurement setup in reflection geometry.}
    \label{fig:thz-tds}

\end{figure}

As a starting point, we used a \gls{thz-tds} platform  (Advantest TAS7500TS), which offers rapid, broadband measurements in reflection or transmission (see, e.g.,~\cite{Grischkowsky:90, Dai:04}). A schematic of the measurement setup~\cite{Vandrevala:19} is shown in Fig.~\ref{fig:thz-tds}.
In this \gls{thz-tds} system, 
a femtosecond pulse from an \gls{ir} laser is incident on a photoconductive antenna, which generates a broadband \gls{thz} pulse. 
After interacting with the sample, the \gls{thz} pulse is detected at a second photoconductive antenna by mixing with a delayed \gls{ir} pulse. 

\begin{figure}
    \centering
    \subfloat[\gls{thz} reflectance vs. metal film \label{fig:thz-tds-spectra-aligned}]{%
    \includegraphics[width=0.90\columnwidth]{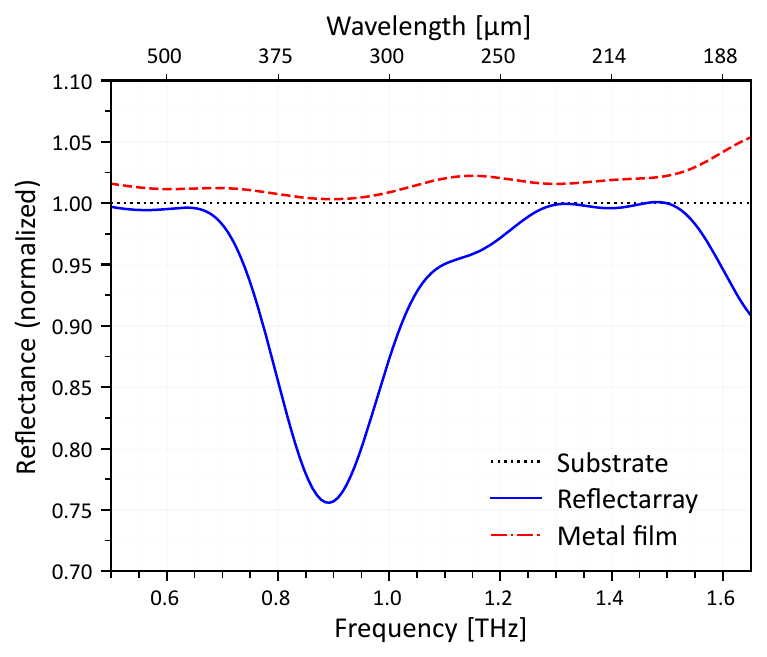}}

    \subfloat[Reflectance at \ang{0} and \ang{90} to \gls{thz} polarization \label{fig:thz-tds-spectra-rotated}]{%
    \includegraphics[width=0.90\columnwidth]{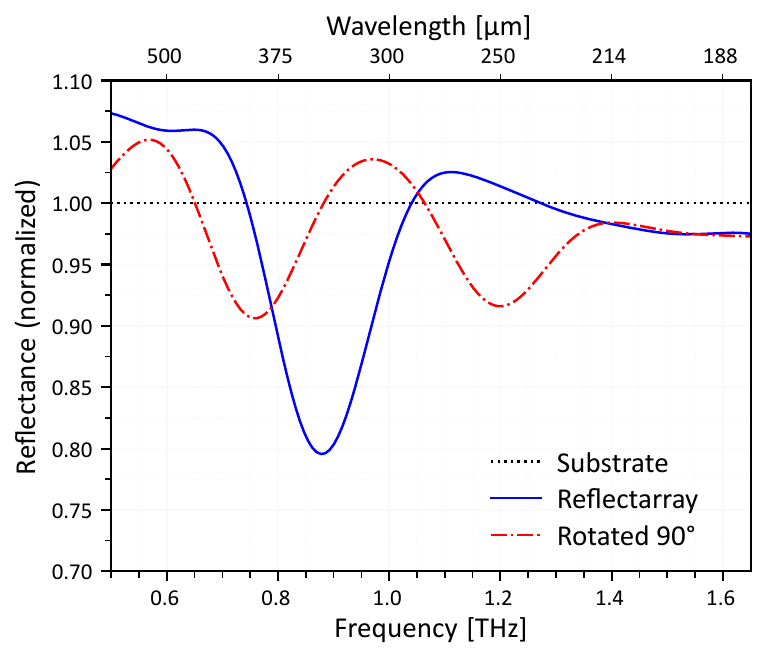}}    
    \caption{\gls{thz-tds} reflectance measurements of the reflectarray normalized to the substrate.}
    \label{fig:thz-tds-spectra}
\end{figure}

The non-specular reflection by the \gls{irs} should direct the component of the broadband \gls{thz} pulse at the resonance frequency away from the detector, resulting in a \textit{decrease} in reflectance. 
This can be clearly seen in the reflectance spectrum plotted in Fig.~\ref{fig:thz-tds-spectra}, where the blue curve shows a clear dip near \SI{1}{\tera\hertz}. 
While designed for \SI{1}{\tera\hertz}, the response is centered at \SI{0.9}{\tera\hertz}, which is related to the tolerance in fabrication, as the stub lengths involved micrometer precision, and the photolithographic process involving complex lift-off had a tolerance of 10\% (\SI{2}{\micro\meter} at \SI{22.5}{\micro\meter} minimum stub length).
The dashed red curve is the reflection of a uniform metal film of equal thickness, and both are normalized to the reflection from the bare substrate. 
The dip at \SI{0.9}{\tera\hertz} is eliminated when the sample is rotated by \ang{90} (Fig.~\ref{fig:thz-tds-spectra-rotated}), confirming that the response is indeed due to the element geometry and not some other factor. 

\subsection{Terahertz Communication Testbed}

We verify the improvement in link quality when utilizing the reflectarray through the \teranova\ testbed~\cite{SEN2020107370}.

\subsubsection{Setup}

The \teranova\ testbed consists of a transmitter with a high-performance analog \gls{psg} and an \gls{awg} up to 92~GSa/s from Keysight Technologies, an upconverter frontend, along with a directional high-gain horn antenna, encompassing true terahertz frequencies (\SIrange{1}{1.05}{\tera\hertz}). The \gls{psg} is used to generate the \gls{lo} signal, which is mixed with the \gls{if} signal generated by the \gls{awg}, and finally upconverted to a higher \gls{rf} signal. 
The upconverter is manufactured by \gls{vdi} and consists of a frequency multiplier chain of $\times 24$, a frequency mixer with a \gls{dsb} conversion loss of \SI{14}{\decibel} and an \gls{if} \gls{lna} with \SI{1}{\decibel} of gain. The transmit power at \gls{rf} before feeding the antenna is about \SI{-12}{dBm} (\SI{60}{\micro\watt}) and the horn antenna gain is \SI{26}{dBi}\@. 
The testbed receiver has a similar design and is equipped with a high-performance \gls{dso} up to 160~GSa/s. 
The downconverter 
has the same architecture as the upconverter but is equipped with a high-gain \gls{if} \gls{lna} providing \SI{12}{\decibel} of gain.

\begin{figure}
    \centering
    \includegraphics[width=\columnwidth]{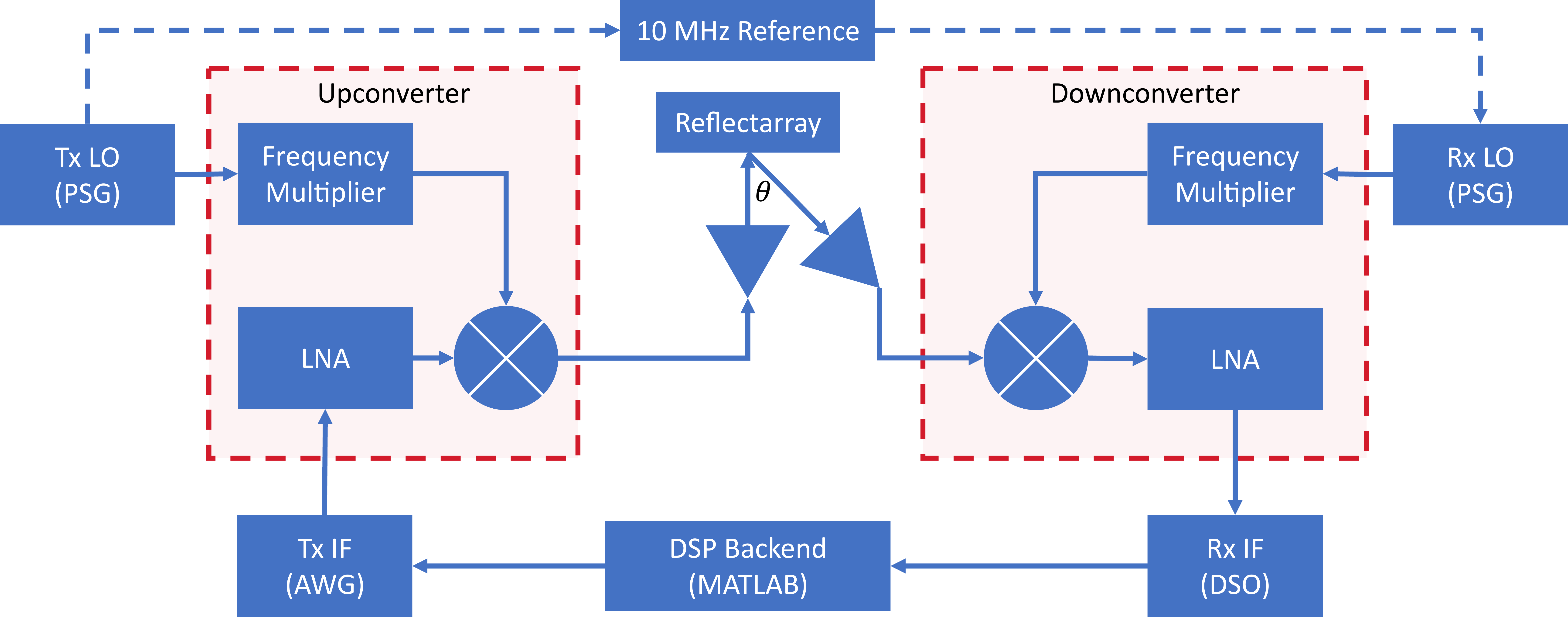}
    \caption{A block diagram depicting the interconnection between the various transmitter and receiver components of the \teranova\ testbed.}
    \label{fig:teranova block diagram}

\end{figure}

Figure~\ref{fig:teranova block diagram} depicts how the different transmitter and receiver components are connected. 
The \SI{10}{\mega\hertz} reference cable is used to synchronize the transmitter and receiver \glspl{psg} and compensate for the carrier frequency and phase offsets.
 

The \gls{dsp} backend of the \teranova\ testbed involves designing discrete-time pulse-shaped modulated waveforms in MATLAB and uploading them to the \gls{awg}\@. The \gls{awg} then generates the modulated \gls{if} signals, which are upconverted to \gls{rf} by the \gls{vdi} upconverters, transmitted over the air, received by the \gls{vdi} downconverters, get downconverted back to \gls{if}, and are finally captured and stored via the \gls{dso}\@. Post-processing and \gls{dsp} techniques can then be applied to the captured signals in MATLAB\@. Given this flexibility, we can experimentally evaluate any arbitrary modulation scheme or \gls{dsp} technique. 

\subsubsection{Controlled Reflection Validation}

\begin{figure}
    \centering
    \includegraphics[width=\columnwidth]{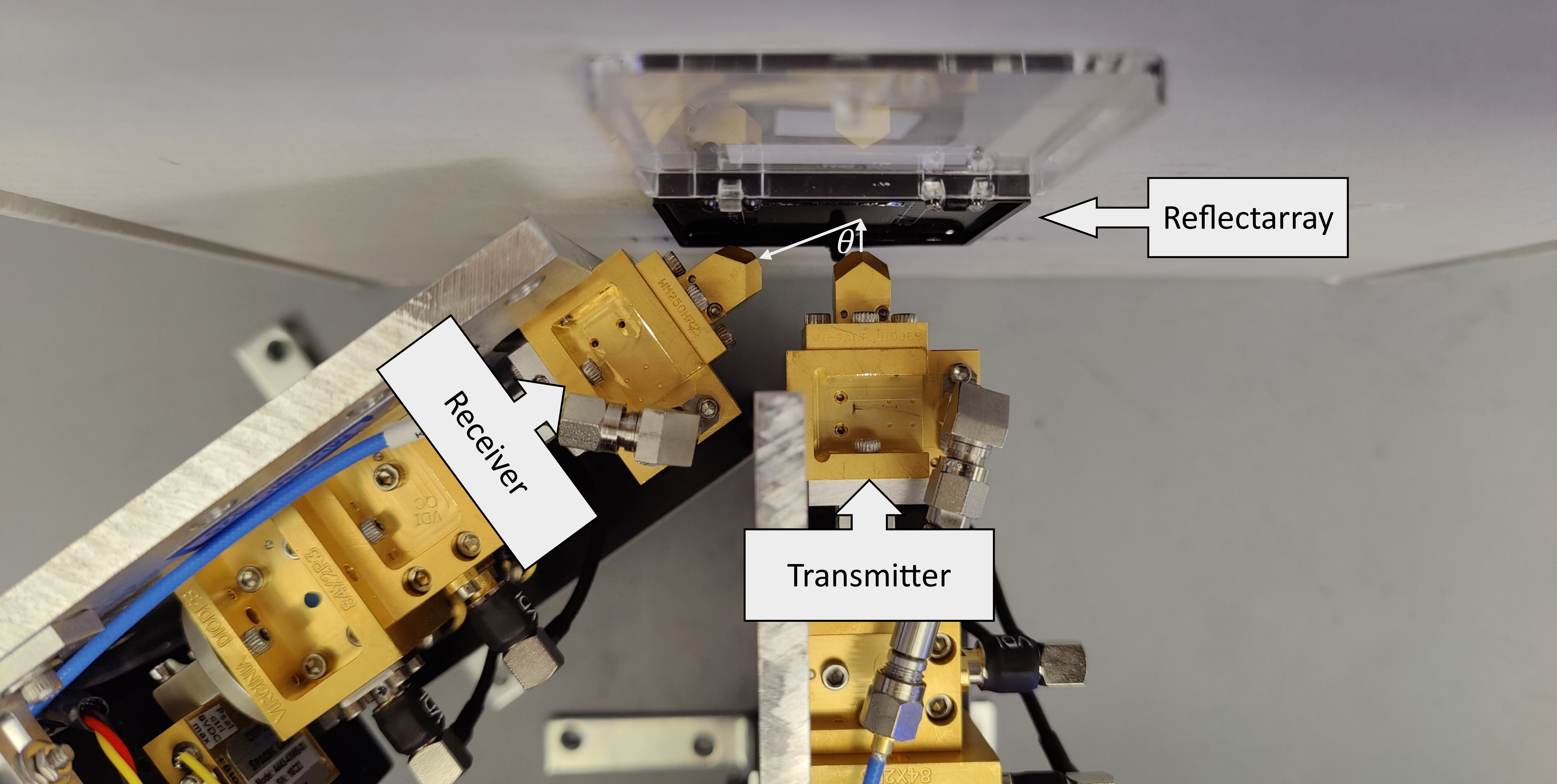}
    \caption{Experimental validation setup of the fabricated metallic reflectarray using the \SIrange{1}{1.05}{\tera\hertz} up and downconverter frontends depicting a controlled reflection scenario at $\theta=\ang{30}$.}
    \label{fig:experimental setup}
\end{figure}

As depicted in Fig.~\ref{fig:experimental setup}, the transmitter is incident normal to the reflectarray, with the receiver placed in the direction of the controlled reflection ($\theta=\ang{30}$). We also replaced the reflectarray with a metallic sheet to act as a benchmark. As a preliminary step, we measure the \gls{snr} when utilizing the fabricated reflectarray both in the case of the receiver being placed at the expected reflection of the signal as well as specular reflection. The \gls{snr} of a \SI{1}{\giga\hertz} \gls{if} sinusoidal signal was found to be \SI{32.9194}{\decibel} in the controlled reflection scenario ($\theta=\ang{30}$) and \SI{1.3899}{\decibel} in the specular reflection scenario (i.e. equal incident and reflected angles from the normal). This clearly indicates that the reflectarray is working as intended and directing the radiation in a non-specular path, potentially enabling \gls{nlos} links.

\begin{figure}
    \centering
    \subfloat[Metallic sheet]{\includegraphics[width=0.90\columnwidth]{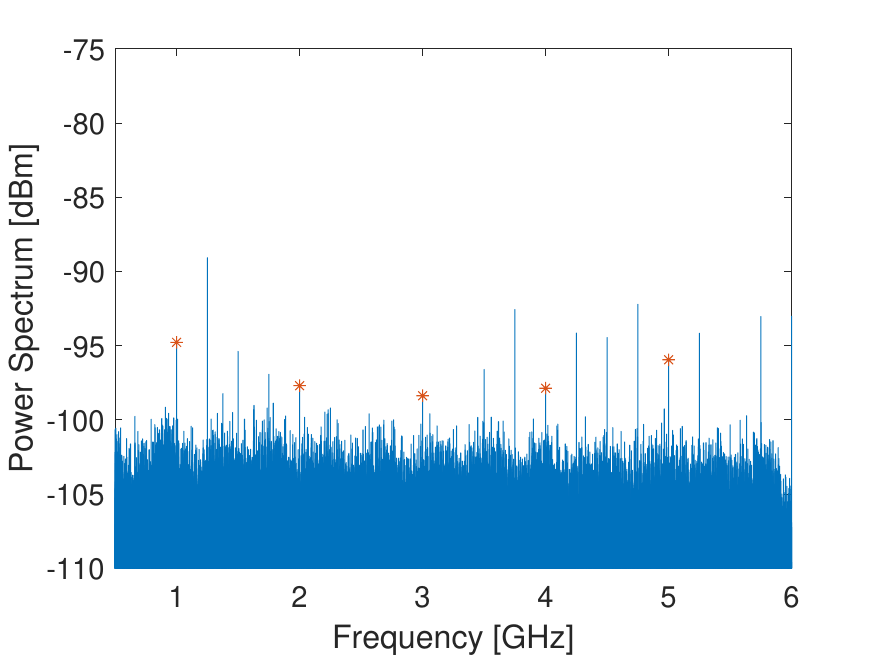}}

    \subfloat[Reflectarray]{\includegraphics[width=0.90\columnwidth]{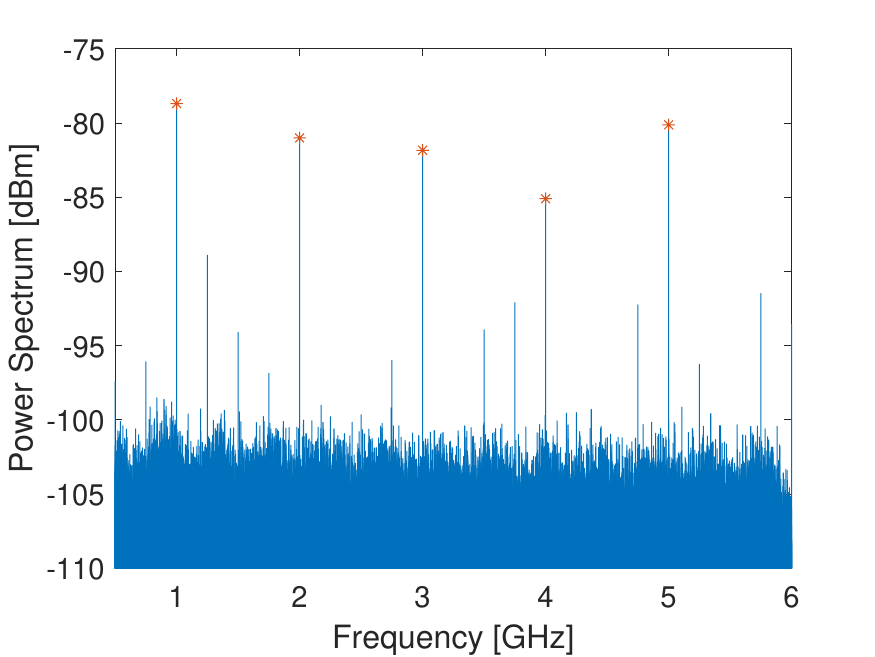}}
    \caption{Power spectrum (\SI{50}{\ohm}) at \gls{if} for the received reflected signal at $\theta=\ang{30}$ with normal incidence.}
    \label{fig:5 tones}
\end{figure}

To further verify broadband operation, we upconvert and transmit an \gls{if} signal comprising the sum of five sinusoidal signals at 1, 2, 3, 4, and \SI{5}{\giga\hertz}. These were successfully received with a high \gls{snr}, as shown in Fig.~\ref{fig:5 tones}, in the case of the reflectarray, but the signals could not be recovered when replacing the reflectarray with a metallic sheet. 

\begin{figure}
    \centering
    \subfloat[Metallic sheet]{\includegraphics[width=0.90\columnwidth]{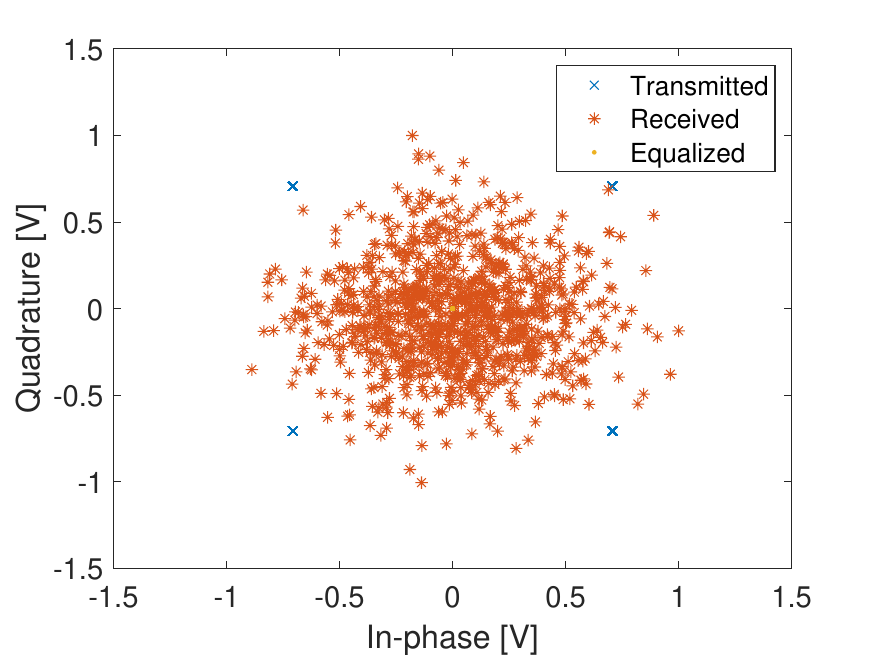}}

    \subfloat[Reflectarray]{\includegraphics[width=0.90\columnwidth]{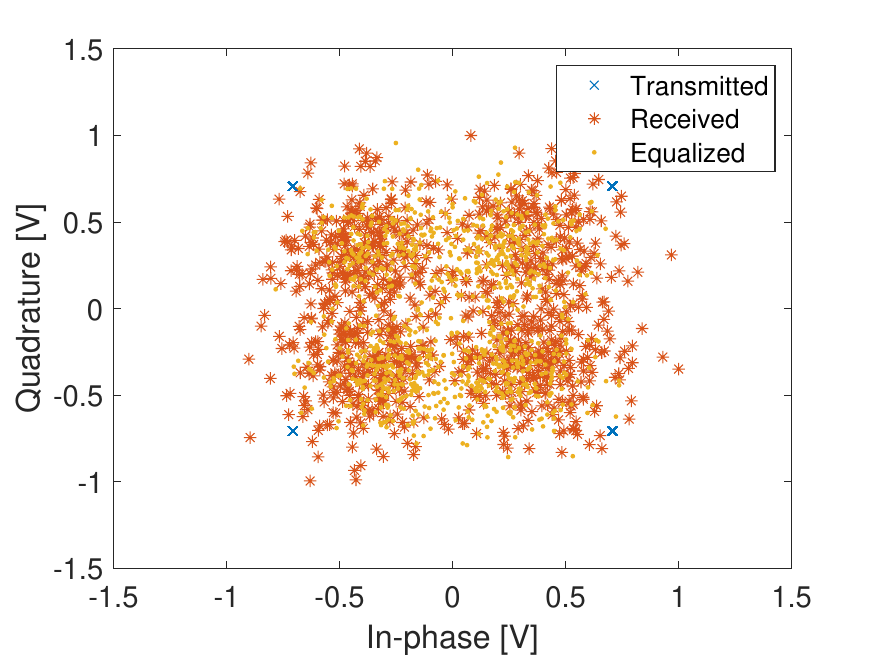}}
    \caption{QPSK constellation diagram (Passband bandwidth = \SI{500}{\mega\hertz}, Bitrate = \SI{500}{Mbps}) for the received reflected signal at $\theta=\ang{30}$ with normal incidence.}
    \label{fig:qpsk constellation}
    
\end{figure}

Next, we transmit a QPSK modulated data signal with a passband bandwidth of \SI{500}{\mega\hertz}, comprising 200~pilot bits and 2000~data bits. The constellation diagram for the transmitted, received, and equalized signals is shown in Fig.~\ref{fig:qpsk constellation}. 
With the reflectarray, the effective received data rate was \SI{495.05}{Mbps}, with a measured \gls{ber} of 0.0335 
without any error correction. This \gls{ber} is very close to the theoretical \gls{awgn} channel \gls{ber} given the received \gls{snr}\@. The measured \gls{ber} when the reflectarray is replaced by a metallic sheet was 0.499, the worst possible \gls{ber}\@. Moreover, the reflectarray can be used for multi-\gls{gbps} data rates and higher modulation orders over multi-gigahertz-wide bandwidths, but it is very challenging to close the link with sufficient \gls{snr} given the very low available transmit power at true terahertz frequencies.
Nonetheless, the results verify the response of the reflectarray and the potential for establishing highly focused and even \gls{nlos} links at \gls{thz} frequencies.  

\section{Conclusion and Future Works}
\label{sec:conclusion}
In this paper, we present 
the first metallic reflectarray achieving \gls{nlos} communications with information-carrying signals at true terahertz frequencies. The results validate the reflect\-array working principle, showing effective communication despite the very low available transmit power. 
In the future, we will work towards increasing the supported bandwidth and communication distance by improving the transmitter power and receiver sensitivity. 
We will also explore the possibility of developing tunable structures by replacing the fixed-length stubs with voltage-controlled delay lines~\cite{singh2020hybrid}.


\begin{acks}
This work was funded by the Air Force Research Laboratory award FA8750-20-1-0500 and the National Science Foundation awards CNS-1955004 and CNS-2011411.
\end{acks}

\bibliographystyle{ACM-Reference-Format}
\bibliography{ref}

\end{document}